\documentstyle[12pt]{article}
\newcommand{\be}{\begin{eqnarray}}
\newcommand{\ee}{\end{eqnarray}}

\begin{document}

\vspace{1cm}

\begin{center}

\LARGE{Electromagnetic form factors of the $\rho$ meson in a light-front
constituent quark model}\\

\vspace{1cm}

\large{ F. Cardarelli$^{(a)}$, I.L. Grach$^{(b)}$, I.M. Narodetskii$^{(b)}$\\
 G. Salm\`{e}$^{(c)}$, S. Simula$^{(c)}$}\\

\vspace{0.5cm}

\normalsize{$^{(a)}$Istituto Nazionale di Fisica Nucleare, Sezione Tor
Vergata\\
Via della Ricerca Scientifica, I-00133 Roma, Italy\\ $^{(b)}$Institute of
Theoretical and Experimental Physics\\ Moscow 117259, Russia\\ $^{(c)}$Istituto
Nazionale di Fisica Nucleare, Sezione Sanit\`{a},\\ Viale Regina Elena 299,
I-00161 Roma, Italy}

\end{center}

\vspace{1cm}

\begin{abstract}

The electromagnetic form factors of the $\rho$ meson are evaluated adopting a
relativistic constituent quark model based on the light-front formalism, and
using a meson wave function with the high-momentum tail generated by the
one-gluon-exchange interaction. The breakdown of the rotational covariance for
the one-body component of the current operator is investigated and the
sensitivity of the ratio of the $\rho$-meson form factors to the pion (charge)
form factor to the spin-dependent component of the effective $q \bar{q}$
interaction is illustrated.

\end{abstract}

\vspace{1cm}

PACS number(s): 12.38.Lg, 12.40.Qq, 12.40.-y, 13.40.Gp, 14.40.Aq, 14.40.Cs

\newpage

\pagestyle{plain}

\indent The understanding of the electroweak properties of hadrons has recently
received much attention within the context of constituent quark models based on
the so-called Hamiltonian light-front formalism \cite{KP91}. As a matter of
fact, light-front quark models have been applied to the evaluation of the
charge form factor of the pion \cite{CAR94,JK90,PION}, the electromagnetic
(e.m.) form factors of the $\rho$ meson \cite{KEI94}, the vector and axial form
factors of the nucleon \cite{NUCLEON}, and the radiative, leptonic and
semileptonic decays of both pseudoscalar and vector mesons \cite{JAUS}. In most
of these applications (\cite{PION} - \cite{JAUS}) it has been assumed that the
hadron wave function is simply given by a harmonic oscillator ansatz, which is
expected to describe the effects of the confinement scale only. However, it has
been shown \cite{CAR94} that the high momentum components generated in the wave
function by the one-gluon-exchange interaction, sharply affect the charge form
factor of the pion for values of the squared four-momentum transfer $Q^2$ up to
few $(GeV/c)^2$. The aim of this letter is to extend to the $\rho$ vector meson
the analysis performed  for the pion in ref. \cite{CAR94}, i.e. to investigate
the sensitivity of the e.m. form factors of $\pi$ and $\rho$ mesons to the
short-range structure of the effective $q \bar{q}$ interaction. The
calculations of the $\rho$-meson form factors presented in this letter, are
based on Poincar\'e-covariant wave functions and one-body e.m. currents. Since
for a spin-1 hadron the rotational covariance of the e.m. current operator is
not ensured by its one-body component alone, the effects of the violation of
the so-called angular condition (see ref. \cite{GK84}) upon the $\rho$-meson
form factors are estimated using wave functions with different high-momentum
tails.

\indent The quark model used in this letter is based on the light-front
formalism which represents the natural framework for constructing a
relativistic
model for the valence $q \bar{q}$ component of a meson. As is known, the
intrinsic light-front kinematical variables are $\vec{k}_{\perp} = \vec{p}_{q
\perp} - \xi \vec{P}_{\perp}$ and $\xi = p^+_q / P^+$, where the subscript
$\perp$ indicates the projection perpendicular to the spin quantization axis,
defined by the vector $\hat{{\bf n}}=(0,0,1)$, and the {\em plus} component of
a 4-vector $p \equiv (p^0, {\bf p} )$ is given by $p^+ = p^0 + \hat{{\bf n}}
\cdot {\bf p}$; eventually, $\vec{P} \equiv (P^+, \vec{P}_{\perp}) = \vec{p}_q
+ \vec{p}_{\bar{q}}$ is the total momentum of the meson. In what follows, only
the $^3S_1$ channel of the $\rho$ meson is considered, being the $D$-wave
component extremely small ($p_D \simeq 0.16 \%$). As a matter of fact, in ref.
\cite{KEI94} it has been checked that a $D$-wave admixture with $p_D \simeq
0.16 \%$ has negligible effects on the e.m. form factors in the $Q^2$-range
considered in this letter. Omitting for the sake of simplicity the flavour and
colour degrees of freedom, the requirement of Poincar\'e covariance for the
intrinsic wave function $\chi^1_{\mu} (\xi, \vec{k}_{\perp}, \nu \bar{\nu})$ of
a $\rho$ meson with helicity $\mu$ implies (cf. ref. \cite{KP91})
 \be
   \chi^1_{\mu}(\xi, \vec{k}_{\perp}, \nu \bar{\nu}) = \sqrt{{M_0 \over 16\pi
   \xi(1 - \xi)}} ~ R_{\mu}(\xi, \vec{k}_{\perp}, \nu \bar{\nu}) ~
w^{\rho}(k^2)
   \label{2}
 \ee
where $\nu, \bar{\nu}$ are the quark spin variables, $k^2 \equiv k^2_{\perp} +
k^2_n$, $k_n \equiv (\xi - 1/2) M_0$, $M_0^2 = (m_q^2 + k^2_{\perp}) / \xi$
$+ ~ (m_{\bar{q}}^2 + k^2_{\perp}) / (1 - \xi)$ and (see ref. \cite{JAUS})
 \be
    R_{\mu} (\xi, \vec{k}_{\perp}, \nu \bar{\nu}) = \sum_{\nu' \bar{\nu'}} ~
    \langle \nu | R_M^{\dag} (\xi, \vec{k}_{\perp}, m_q) | \nu' \rangle ~
    \langle \bar{\nu} | R_M^{\dag} (1 - \xi, - \vec{k}_{\perp}, m_{\bar{q}}) |
    \bar{\nu'} \rangle ~ \langle {1 \over 2} \nu' {1 \over 2} \bar{\nu'} | 1
    \mu \rangle
    \label{3}
 \ee
with $m_q$ ($m_{\bar{q}}$) being the constituent quark (antiquark) mass and
$R_M$ the $2 \times 2$ irreducible representation of the Melosh rotation
\cite{MEL74}.

\indent As in ref. \cite{CAR94}, the radial wave function $w^{\rho}(k^2)$
appearing in Eq. (\ref{2}) is identified with the equal-time radial wave
function in the $\rho$-meson rest-frame. In this letter we will adopt the
effective $q \bar{q}$ Hamiltonian introduced by Godfrey and Isgur (GI)
\cite{GI85} for reproducing the meson mass spectra, viz.
 \be
    H_{q \bar{q}} ~ w^{q \bar{q}}(k^2) | j \mu \rangle & \equiv & \left
    [\sqrt{m_q^2 + k^2} + \sqrt{m_{\bar{q}}^2 + k^2} + V_{q \bar{q}} \right ] ~
    w^{q \bar{q}}(k^2) | j \mu \rangle  =  M_{q \bar{q}} w^{q \bar{q}}(k^2)
    | j \mu \rangle
    \label{7}
 \ee
where $M_{q \bar{q}}$ is the mass of the meson, $| j \mu \rangle = \sum_{\nu
\bar{\nu}} \langle {1 \over 2} \nu {1 \over 2} \bar{\nu} | j \mu \rangle
\chi_{\nu} \chi_{\bar{\nu}}$ is the equal-time quark-spin wave function and
$V_{q \bar{q}}$ is the effective $q \bar{q}$  potential. The interaction in the
GI scheme, $V_{(GI)}$, is composed by a one-gluon-exchange (OGE) term (dominant
at short separations) and a linear-confining term (dominant at large
separations). In order to analyze the effects of different components of the GI
interaction, two other choices of $w^{q \bar{q}}(k^2)$ will be considered; the
first one is the solution of Eq. (\ref{7}) obtained after switching off the OGE
part of $V_{(GI)}$, i.e., by retaining only its linear confining term,
$V_{(conf)}$, whereas the second choice is given by the solution of Eq.
(\ref{7}) obtained when only the spin-independent part, $V_{(si)}$, of
$V_{(GI)}$ is considered. The three different forms of $w^{q \bar{q}}(k^2)$
will be denoted hereafter by $w^{q \bar{q}}_{(conf)}$, $w^{q \bar{q}}_{(si)}$
and  $w^{q \bar{q}}_{(GI)}$ corresponding to $V_{(conf)}$, $V_{(si)}$ and
$V_{(GI)}$, respectively. It should be pointed out that the pion ($^1S_0$
channel) and $\rho$-meson ($^3S_1$ channel) radial wave functions differ only
when the spin-spin component of the $q \bar{q}$ interaction is considered; this
means that: $w^{\pi}_{(conf)} = w^{\rho}_{(conf)} \equiv w_{(conf)}$,
$w^{\pi}_{(si)} = w^{\rho}_{(si)} \equiv w_{(si)}$ and $w^{\pi}_{(GI)} \neq
w^{\rho}_{(GI)}$. The four wave functions $w_{(conf)}$, $w_{(si)}$,
$w^{\pi}_{(GI)}$ and $w^{\rho}_{(GI)}$ are shown in fig. 1. It can clearly be
seen that both the central and the spin-dependent components of the OGE
interaction strongly affect the high-momentum tail of the $\pi$- and
$\rho$-meson wave functions. It should also be reminded that the radial wave
function $w_{(conf)}$ turns out to be very close to the simple harmonic
oscillator ansatz adopted in many light-front calculations (see ref.
\cite{CAR94}). According to ref. \cite{GI85}, the value $m_q = m_{\bar{q}} =
0.220~GeV$ is adopted.

\indent{\bf Matrix elements of the electromagnetic current.} Within the
light-front formalism (cf. ref. \cite{CCKP88}), all the invariant form factors
of a hadron can be determined using only the matrix elements of the component
$I^+(0)$ of the current operator evaluated in an appropriate frame, which, for
spacelike four-momentum transfer $Q^2$, can be identified with the Breit frame
where $Q^+ = 0$, $P^+ = P'^+ = \sqrt{M^2_{\rho} + Q^2 / 4}$ and
$\vec{P'}_{\perp} = -\vec{P}_{\perp} = \vec{Q}_{\perp} / 2$. In the case of a
spin-1 hadron only three independent (invariant) form factors exist, whereas
four matrix elements of $I^+(0)$ are independent after considering the
properties of $I^+(0)$ to be Hermitean and invariant under i) time
reversal, ii) rotations about $\hat{{\bf n}}$, and iii)  reflection on the
plane
perpendicular to $\hat{{\bf n}}$. An additional condition comes from the
rotational invariance of the charge density, which involves unitary
transformations based upon a subset of Poincar\'e generators depending on the
interaction. Thus, the additional constraint, usually called the angular
condition, is not generally satisfied by the matrix elements of the one-body
current alone, and requires the existence of many-body currents. The angular
condition can be written in the following form \cite{GK84}
 \be
    \Delta(Q^2) \equiv (1 + 2 \eta) I_{11} +I_{1-1} - \sqrt{8 \eta} I_{10} -
    I_{00} = 0
    \label{8}
 \ee
where $I_{\mu' \mu} \equiv <\vec{P'} \mu'| I^+(0) |\vec{P} \mu>$ stands for the
matrix elements of $I^+(0)$ in the Breit frame and  $\eta \equiv Q^2 / 4
M^2_{\rho}$, with $M_{\rho} = 0.77~GeV$ being the experimental $\rho$-meson
mass. In ref. \cite{CCKP88} the following relations between the matrix elements
$I_{\mu' \mu}$ and the (conventional) invariant form factors $G_0$, $G_1$ and
$G_2$, have been obtained
 \be
   [G_0]^{CCKP}& = & {1 \over 3(1 + \eta)} [({3 \over 2} - \eta) (I_{11}
                   + I_{00}) + 5\sqrt{2\eta} I_{10} + (2\eta - {1 \over 2})
                   I_{1-1}] \nonumber    \\
   {[G_1]}^{CCKP}& = &{1 \over 1 + \eta} [I_{11} + I_{00} - I_{1-1} - {2(1 -
\eta)
                   \over \sqrt{2\eta}} I_{10}] \nonumber \\
   {[G_2]}^{CCKP}& = &{\sqrt{2} \over 3(1 + \eta)} [-\eta I_{11} +
2\sqrt{2\eta}
                   I_{10} - \eta I_{00} - (\eta + 2) I_{1-1}]
   \label{9}
 \ee
In ref. \cite{FFS93} $G_0$ has been obtained using a specific criterion for
choosing the "good" matrix elements of $I^+(0)$ in the infinite momentum frame
and in the Breit one, viz.
 \be
    [G_0]^{FFS} = {1 \over 3(1 + \eta)} [(2\eta + 3) I_{11} + 2\sqrt{2\eta}
                  I_{10} - \eta I_{00} + (2\eta + 1) I_{1-1}]
                  \nonumber \\
    {[G_1]}^{FFS} = [G_1]^{CCKP} ~~~~, ~~~~ {[G_2]}^{FFS} = [G_2]^{CCKP}
    \label{10}
 \ee
In ref. \cite{GK84} the "worst" matrix element has been assumed to be $I_{00}$.
After eliminating $I_{00}$ from Eq. (\ref{9}) through the angular condition
(\ref{8}), one has
 \be
    [G_0]^{GK}& = & {1 \over 3}[(3 - 2\eta) I_{11}+ 2\sqrt{2\eta} I_{10}
                    + I_{1-1}] \nonumber \\
    {[G_1]}^{GK}& = & 2[I_{11} - {1 \over \sqrt{2\eta}} I_{10}] \nonumber \\
    {[G_2]}^{GK}& = & {2\sqrt{2} \over 3} [-\eta I_{11} + \sqrt{2\eta} I_{10}
                       - I_{1-1}]
    \label{11}
 \ee
Following refs. \cite{BH92} and \cite{LB80}, the matrix element $I_{00}$ is
expected to be the dominant one in the perturbative QCD regime; if the matrix
element $I_{11}$, instead of $I_{00}$, is eliminated from Eq. (\ref{9}) through
the angular condition (\ref{8}), one gets
 \be
    [G_0]^{BH}& = &{1 \over 3 (1 + 2 \eta)} [(3 - 2 \eta) I_{00} +
                 8 \sqrt{2 \eta} I_{10} + 2 (2 \eta - 1) I_{1-1}] \nonumber \\
    {[G_1]}^{BH}& = &{2 \over 1 + 2 \eta} [I_{00} - I_{1-1} + (2 \eta - 1)
                 {I_{10} \over \sqrt{2 \eta}}] \nonumber \\
    {[G_2]}^{BH}& = &{2 \sqrt{2} \over 3(1 + 2 \eta)} [\sqrt{2 \eta} I_{10} -
                 \eta I_{00} - (1 + \eta) I_{1-1}]
    \label{12}
 \ee
It should be pointed out that, if the exact Poincar\'e-covariant (many-body)
$I^+(0)$ current is used, all the prescriptions (like those specified by Eqs.
(\ref{9})-(\ref{12})) should yield the same results for the invariant form
factors $G_i$, whereas, when only the one-body component of the e.m. current
operator is considered, the angular condition (\ref{8}) is in general violated
(i.e., $\Delta(Q^2) \neq 0$) and the calculation of the $G_i$ depends upon the
prescription used.

\indent As for the one-body part of the e.m. current operator, the expression
$I^+(0) = $ $\sum_{i = q, \bar{q}} $ $e_i ~ F^i(Q^2) $ $\gamma_i^+$, where
$F^i(Q^2)$ is the charge form factor of the constituent quark, has been
adopted; indeed, according to the findings of refs. \cite{NUCLEON}(a) and
\cite{KUKD}, the anomalous magnetic moments of the constituent quarks are
expected to be small and, therefore, only Dirac magnetic moments are considered
in this letter. Thus, using Eq. (\ref{2}), the matrix elements $I_{\mu' \mu}$
appearing in Eqs. (\ref{9}) - (\ref{12}) can be written as
 \be
    I_{\mu' \mu} = F(Q^2) \int d \xi d \vec{k}_{\perp}{\sqrt{M'_0 M_0}
    \over 16 \pi \xi(1 - \xi)} {\cal M}_{\mu' \mu} (\xi, \vec{k}_{\perp},
    \vec{k'}_{\perp}) w^{\rho}(k'^2) w^{\rho}(k^2)
    \label{13}
 \ee
where $\vec{k'}_{\perp} \equiv \vec{k}_{\perp} + (1 - \xi) \vec{Q}_{\perp}$,
$F(Q^2) = e_q F^q(Q^2) + e_{\bar{q}} F^{\bar{q}}(Q^2)$ and ${\cal M}_{\mu'\mu}
\equiv \sum_{\nu \bar{\nu}}$ $R_{\mu} (\xi, \vec{k}_{\perp}, \nu \bar{\nu})$
$R_{\mu'}^* (\xi, \vec{k}'_{\perp}, \nu \bar{\nu})$ arise from the Melosh
rotation  of the quark spins. Note that both the use of $I^+(0)$ and the
choice $Q^+ = 0$ allow to suppress the contribution of the so-called Z-graph
(pair creation from the vacuum) \cite{LB80,ZGRAPH}.

\indent {\bf Results of calculations.} The invariant form factors $G_i$ have
been evaluated using the CCKP (Eq. (\ref{9})), FFS (Eq. (\ref{10})), GK
(Eq. (\ref{11})) and BH (Eq. (\ref{12})) prescriptions. As already pointed
out, any dependence of the calculations upon the prescription used is a
consequence of the breakdown of the angular condition (Eq. (\ref{8})), which is
directly expressed by the departure of the quantity $\Delta(Q^2)$ from zero. In
refs. \cite{GK84} and \cite{FFS93} it has been found that the effects of the
violation of the angular condition upon the form factors of the deuteron is
small at all accessible values of $Q^2$, though $\Delta(Q^2)$ turns out to be
an increasing function of $Q^2$. In the case of the $\rho$ meson, for which the
momentum of the constituent is not small with respect to its mass (see fig. 1),
the breakdown of the angular condition is expected to have large effects on the
calculated form factors \cite{KEI94}. By adopting for the radial wave function
$w^{\rho}$ the choices $w_{(conf)}$, $w_{(si)}$ and $w^{\rho}_{(GI)}$ and
neglecting the charge form factor of the constituent quarks (i.e., assuming
$F(Q^2) = 1$ in Eq. (\ref{13})), the matrix elements $I_{\mu' \mu}$ have been
calculated and the results obtained for the quantity $\Delta(Q^2)$ are reported
in fig. 2. It can be seen that the violation of the angular condition is
strongly affected by the high-momentum tail of the $\rho$-meson wave function;
this means that the two-body currents required to restore the rotational
covariance of the e.m. current operator are expected to be sharply sensitive to
the short-range structure of the effective $q \bar{q}$ interaction.

\indent Using the wave function $w^{\rho} _{(GI)}$ and assuming $F(Q^2) = 1$
in Eq. (\ref{13}), the sensitivity of the form factors $G_i(Q^2)$ to the
prescriptions given by Eqs. (\ref{9}) - (\ref{12}) is illustrated in fig. 3. It
can be seen that all the form factors are sensitive to the prescription used
only for $Q^2 \geq 0.5~(GeV/c)^2$; in particular, $G_2 (Q^2)$ is strongly
affected by the violation of the angular condition, so that the difference from
its non-relativistic limit (i.e. $G_2(Q^2) = 0$ if the $D$-wave is disregarded)
might be significantly reduced. In agreement with the findings of ref.
\cite{KEI94}, where a soft wave function was adopted, the charge radius of the
$\rho$ meson ($<r^2> \equiv lim_{Q^2 \rightarrow 0} ~ 6 (1 - G_0(Q^2)) / Q^2$)
is slightly affected by the prescription used also in the case of wave
functions with a high-momentum tail. In particular, a spread of $\sim 10-15\%$
around the value $<r^2> = 0.35~fm^2$, calculated using the CCKP  prescription
and assuming $F(Q^2) = 1$, has been obtained. Moreover, we have found that the
values of the magnetic ($\mu_1 \equiv lim_{Q^2 \rightarrow 0} ~ G_1(Q^2)$) and
quadrupole ($Q_2 \equiv lim_{Q^2 \rightarrow 0} ~ 3 \sqrt{2} G_2(Q^2) / Q^2$)
moments, which are independent of the violation of the angular condition (cf.
ref. \cite{KS84}), are $\mu_1 = 2.26$ (i.e., $\sim 10\%$ larger than its
non-relativistic value ($\mu_1 = 2$)) and $Q_2 = 0.024~fm^2$, respectively.

\indent Let us now investigate the effects of the different components of the
GI
interaction upon the form factors $G_i (Q^2)$. In order to get rid of
contributions arising from the charge form factor of the constituent quarks
($F(Q^2)$ in Eq. (\ref{13})) as well as to enhance the sensitivity of the
calculations to the spin-spin component of the effective $q \bar{q}$
interaction, it is convenient to compare the form factors $G_i(Q^2)$ of the
$\rho$-meson ($^3S_1$ channel) with the charge form factor $F_{\pi}(Q^2)$ of
the pion ($^1S_0$ channel) by considering the ratios $R_i(Q^2) \equiv G_i (Q^2)
/ F_{\pi} (Q^2)$, where $F_{\pi} (Q^2)$ is explicitely given by (cf., e.g.,
ref.
\cite{CAR94})
 \be
    F_{\pi}(Q^2) = F(Q^2) \int d\vec{k}_{\perp} d\xi ~ {\sqrt{M_0 M'_0} \over
   16 \pi \xi (1 - \xi)} ~ {\xi (1 - \xi) M_0^2 + \vec{k}_{\perp} \cdot
   \vec{Q}_{\perp} \over  \xi (1 - \xi) M_0 M'_0} ~ w^{\pi}(k'^2) w^{\pi}(k^2)
    \label{18}
 \ee
The radial wave functions of the $\rho$ and $\pi$ mesons corresponding to
the interactions $V_{(conf)}$, $V_{(si)}$ and $V_{(GI)}$ have been considered
in
Eqs. (\ref{13}) and (\ref{18}), respectively. The results of the calculations,
obtained using the CCKP prescription (Eq. (\ref{9})) for the $\rho$ meson, are
reported in fig. 4. It should be pointed out that similar results can be
obtained using the FFS, GK or BH prescriptions instead of the CCKP one. From
fig. 4 it can be seen that: ~ i) both at low and moderate values of $Q^2$ the
ratios $R_i (Q^2)$ are strongly affected by the high momentum components
generated in the meson wave functions by the spin-dependent part of the GI
effective $q \bar{q}$ interaction; ~ ii) for $Q^2 \geq 0.5~(GeV/c)^2$ the
sensitivity to the high momentum tail appears to be of the same order of
magnitude of the uncertainties related to the violation of the angular
condition (cf. fig. 3).

\indent In conclusion, the e.m. form factors of the $\rho$ meson have been
evaluated within a relativistic constituent quark model based on the
light-front formalism, for values of $Q^2$ up to few $(GeV/c)^2$. The effects
of the breakdown of the rotational covariance of the one-body e.m. current
operator as well as the sensitivity of the calculations to the high momentum
tail of the meson wave functions, generated by the one-gluon-exchange
interaction, have been investigated. The main results of our analysis can be
summarized as follows: the ratio of the $\rho$-meson form factors to the pion
(charge) form factor is remarkably sensitive to the high-momentum components
of the meson wave function and, therefore, could allow to investigate the
spin-dependent part of the effective $q \bar{q}$ interaction; however, at
$Q^2 > 0.5~(GeV/c)^2$, such a sensitivity is partially hindered by the
theoretical uncertainties related to the effects of two-body currents, which
are required to ensure the full Poincar\'e covariance of the e.m. current
operator. Therefore, the evaluation of the effects of the (interaction
dependent) two-body currents upon meson form factors is mandatory in order to
get quantitative information on the short-range structure of the meson wave
functions.

\vspace{0.5cm}

We gratefully acknowledge S. Brodsky, B.D. Keister, F. Lev, E. Pace and M.I.
Strikman for many enlightening discussions.

\newpage

\vspace{0.5cm}

\begin{center}

{\bf Figure Captions}

\end{center}

\vspace{0.5cm}

Fig. 1. Wave functions $(k \cdot w^{\pi})^2$ and $(k \cdot w^{\rho})^2$,
calculated using in Eq. (\ref{7}) different effective $q \bar{q}$ interactions,
as a function of the relative momentum $k$. Dotted line: $w^{\pi} = w^{\rho} =
w_{(conf)}$, corresponding to the case in which only the linear confining part
of the GI $q \bar{q}$ interaction \cite{GI85} is considered. Dashed line:
$w^{\pi} = w^{\rho} = w_{(si)}$, corresponding to the solution of Eq. (\ref{7})
obtained using the spin-independent part of the GI interaction. The solid and
dot-dashed lines correspond to  $w^{\rho} = w^{\rho}_{(GI)}$ and $w^{\pi} =
w^{\pi}_{(GI)}$, respectively, obtained by including in Eq. (\ref{7}) the full
spin-dependent GI interaction.

\vspace{0.5cm}

Fig. 2. The quantity $\Delta (Q^2)$ (see Eq. (\ref{8})) as a function of $Q^2$
calculated using various choices of the wave function $w^{\rho}$ appearing in
Eq. (\ref{13}). The dotted, dashed and solid lines correspond to $w^{\rho} =
w_{(conf)}$, $w^{\rho} = w_{(si)}$ and $w^{\rho} = w^{\rho}_{(GI)}$,
respectively (see fig. 1). Calculations have been performed assuming $F(Q^2) =
1$ in Eq. (\ref{13}).

\vspace{0.5cm}

Fig. 3. The invariant form factors $G_i (Q^2)$ of the $\rho$ meson as a
function of $Q^2$ calculated within various prescriptions. The solid, dashed,
dotted and dot-dashed lines correspond to the CCKP (Eq. (\ref{9})), FFS (Eq.
(\ref{10})), GK (Eq. (\ref{11})) and BH (Eq. (\ref{12})) prescriptions,
respectively. Calculations have been performed using $w^{\rho} =
w^{\rho}_{(GI)}$ in Eq. (\ref{13}), which corresponds to the ground-state wave
function of the full GI Hamiltonian \cite{GI85} for the $^3S_1$ channel. In
all the calculations the constituent quark form factor has been neglected
(i.e.,
$F(Q^2) = 1$ in Eq. (\ref{13})). Note that, as for $G_1 (Q^2)$ and $G_2 (Q^2)$,
the FFS prescription coincides with the CCKP one (cf. Eq. (\ref{10})).

\vspace{0.5cm}

Fig. 4. The ratios $R_i (Q^2) \equiv G_i (Q^2) / F_{\pi} (Q^2)$ as a function
of
$Q^2$, calculated using various choices of the $\rho$- and $\pi$-meson wave
functions appearing in Eqs. (\ref{13}) and (\ref{18}), respectively.
Calculations for the $\rho$ meson have been performed using the CCKP
prescription (Eq. (\ref{9})). Dotted line: $w^{\pi} = w^{\rho} = w_{(conf)}$,
corresponding to the case in which only the linear confining part of the GI $q
\bar{q}$ interaction \cite{GI85} is considered in Eq. (\ref{7}). Dashed line:
$w^{\pi} = w^{\rho} = w_{(si)}$, corresponding to the solution of Eq. (\ref{7})
obtained using the spin-independent part of the GI interaction. Solid line:
$w^{\pi} = w^{\pi}_{(GI)}$ and $w^{\rho} = w^{\rho}_{(GI)}$, obtained by
including in Eq. (\ref{7}) the full spin-dependent GI interaction.

\end{document}